\documentclass[useAMS,usenatbib,onecolumn]{mnras}
\usepackage{amsmath,bm,amssymb}
\usepackage{dcolumn}
\usepackage[utf8]{inputenc}
\usepackage{graphics,graphicx}
\usepackage{float}
\usepackage{threeparttable}
\usepackage{url}
\usepackage{epstopdf}
\usepackage{soul}

\usepackage{booktabs}
\usepackage{multirow}

\usepackage{color}

\newcommand{\bra}[1]{\langle #1|}
\newcommand{\ket}[1]{|#1\rangle}

\newcommand{\X}{$\tilde{X}\,^2\Sigma^+$}
\newcommand{\A}{$\tilde{A}\,^2\Pi$}
\newcommand{\B}{$\tilde{B}\,^2\Sigma^+$}

\def\pdu#1#2{\frac{\partial#1}{\partial#2}}

\newcommand{\EVEREST}{{\sc EVEREST}}

\title[ExoMol line lists -- XLVII. CaOH]{ExoMol line lists -- XLVII. Rovibronic molecular line list of the calcium monohydroxide radical (CaOH)}

\date{\today}

\author[A. Owens et al.]
{Alec Owens$^{1}$\thanks{The corresponding author: alec.owens.13@ucl.ac.uk}, Alexander Mitrushchenkov$^{2}$ \thanks{The corresponding author: Alexander.Mitrushchenkov@univ-eiffel.fr}, Sergei N. Yurchenko$^{1}$\thanks{The corresponding author: s.yurchenko@ucl.ac.uk} and Jonathan Tennyson$^{1}$\thanks{The corresponding author: j.tennyson@ucl.ac.uk}\vspace*{4mm}\\
$^1$ Department of Physics and Astronomy, University College London, Gower Street, WC1E 6BT London, UK\\
$^2$ MSME, Universit\'{e} Gustave Eiffel, CNRS UMR 8208, Univ Paris Est Creteil, F-77474 Marne-la- Vallée, France}

\date{Accepted XXXX. Received XXXX; in original form XXXX}

\pagerange{\pageref{firstpage}--\pageref{lastpage}} \pubyear{2021}

\begin{document}

\label{firstpage}

\maketitle

\begin{abstract}
Any future detection of the calcium monohydroxide radical (CaOH) in stellar and exoplanetary atmospheres will rely on accurate molecular opacity data. Here, we present the first comprehensive molecular line list of CaOH covering the \A--\X\ rotation-vibration-electronic and \X--\X\ rotation-vibration bands. The newly computed OYT6 line list contains over 24.2 billion transitions between 3.2 million energy levels with rotational excitation up to $J=175.5$. It is applicable to temperatures up to $T=3000$~K and covers the 0\,--\,35\,000~cm$^{-1}$ range (wavelengths $\lambda > 0.29$~$\mu$m) for rotational, rotation-vibration and the \A--\X\ electronic transition. The strong band around 16\,000~cm$^{-1}$ ($\lambda = 0.63$~$\mu$m) is likely to be of interest in future astronomical observations, particularly in hot rocky exoplanets where temperatures can become extremely high. The OYT6 line list has been generated using empirically-refined \X\ and \A\ state potential energy surfaces, high-level \textit{ab initio} transition dipole moment surfaces and a rigorous treatment of both Renner-Teller and spin-orbit coupling effects, which are necessary for correctly modelling the CaOH spectrum. Post-processing of the CaOH line list has been performed so as to tailor it to high-resolution applications, i.e.\ by replacing calculated energy levels with more accurate empirically-derived values (where available), hence improving the accuracy of the predicted line positions in certain regions. The OYT6 line list is available from the ExoMol database at \href{http://www.exomol.com}{www.exomol.com} and the CDS astronomical database.
\end{abstract}


\section{Introduction}


The calcium monohydroxide radical ($^{40}$Ca$^{16}$O$^{1}$H) has been cited as a missing opacity source in the study of M-dwarf photospheres using the BT-Settl model~\citep{13RaReAl.CaOH}. The atmospheres of hot rocky super-Earth exoplanets are also expected to contain spectroscopic signatures of CaOH~\citep{09Bernath.exo,jt693} but a lack of molecular line list data in the necessary spectral regions inhibits  detections. Notably, it is the rotation-vibration-electronic (rovibronic) spectrum composed of the strong \A--\X\ band system that is of interest. This has motivated the ExoMol database~\citep{jt810,jt631,jt528} to work on the production of a high-accuracy line list for CaOH and this paper is the final stage in that process.

In previous work, we performed an extensive literature search and extracted all meaningful rovibronic spectroscopic transitions (around 3200) of CaOH~\citep{jt791} and processed them using the MARVEL (measured active rotation vibration energy levels) algorithm~\citep{jt412,07CsCzFu.method,12FuCsxx.methods,jt750}. This procedure produced a consistent dataset of empirical-quality energy levels, each possessing a measurement uncertainty and unique quantum number labelling. The advantages of having such a dataset are twofold: Firstly, the theoretical spectroscopic model of the molecule, primarily the molecular potential energy surface (PES), can be fine-tuned in calculations so as to improve the accuracy of the predicted line positions and to a lesser extent the line intensities. Secondly, when post-processing the final line list the computed energy levels can be substituted with the more accurate MARVEL values (where available). Doing so can dramatically improve the accuracy of certain line positions, making the final list list suitable in certain windows for studying exoplanet atmospheres at high spectral resolution~\citep{14Snellen,18Birkby}. Currently, the adaptation of molecular line lists in the ExoMol database for high-resolution applications is being actively pursued, for example, see \citet{jt835}.

A full-dimensional \textit{ab initio} spectroscopic model of the \A--\X\ band system of CaOH has also been reported by the authors~\citep{jt838}. High-level \textit{ab initio} theory was used to construct new potential energy and transition dipole moment surfaces, and both Renner-Teller and spin-orbit coupling effects, which are essential for correctly reproducing the CaOH spectrum, were accounted for in the calculations. Whilst the ground \X\ potential energy surface (PES) was empirically refined to the MARVEL dataset of energy levels, vastly improving its accuracy, the \A\ state PES was not rigorously refined as further development of the computer program \EVEREST~\citep{EVEREST} was necessary. This has now been completed for the present work, ultimately leading to considerably more accurate \A\ state energy levels and $\tilde{A}$--$\tilde{X}$ transition wavenumbers (up to several orders-of-magnitude). Doing so has paved the way for the production of a comprehensive, high-temperature line list of CaOH which we report below.

It is worth mentioning that the spectroscopic model and associated $\tilde{A}$--$\tilde{X}$ line list produced in this work could greatly assist the design of laser cooling schemes in ultracold molecule research and precision tests of fundamental physics. These fields have been focusing on the alkaline earth monohydroxide radicals, notably CaOH~\citep{19KoStYu.CaOH,19AuBoxx.CaOH}, because of the favourable rovibronic energy level structure. Already, direct laser cooling of CaOH to temperatures near 1~millikelvin has been performed in a one-dimensional magneto-optical trap~\citep{20BaViHa.CaOH} and robust experimental schemes to extend cooling of CaOH into the microkelvin regime have been suggested~\citep{21BaViHa.CaOH} and successfully implemented~\citep{22ViHaAn.CaOH}. The calculation of Einstein $A$ coefficients between the different molecular states in CaOH would allow branching ratios and decay routes to be analysed, and such work has previously been carried out for other molecules using ExoMol line lists~\citep{jt663}.

\section{Methods}

\subsection{Refinement of the \A\ potential energy surface}

In CaOH, the Renner-Teller effect~\citep{Renner1934} causes the \A\ state to split into two Renner surfaces $A^{\prime}$ and $A^{\prime\prime}$ at bent molecular geometries. Any correct description of the spectrum of CaOH  must account for the Renner-Teller effect~\citep{96LiCoxx.CaOH,95LiCoxx.CaOH,94CoLiPr.CaOH,92LiCoxx.CaOH,92JaBexx.CaOH,91CoLiPr.CaOH,83HiQiHa.CaOH}. In our spectroscopic model, the $A^{\prime}$ and $A^{\prime\prime}$ surfaces are each represented by their own analytic function, essentially an eighth-order polynomial in terms of linear expansion variables of the bond lengths and bond angle, see \citet{jt838} for a detailed description. The $A^{\prime}$ and $A^{\prime\prime}$ potential functions share the same parameter values for stretching only terms but possess different parameter values for terms involving the bending mode.

Despite numerous studies of the $\tilde{A}$--$\tilde{X}$ band system of CaOH~\citep{85BeBrxx.CaOH,91CoLiPr.CaOH,92LiCoxx.CaOH,92CoLiPr.CaOH,94CoLiPr.CaOH,95LiCoxx.CaOH,06DiShWa.CaOH}, only four vibrational levels of the \A\ state have been characterised experimentally, namely the ground vibrational state $(0,0,0)$, the first excited Ca--O stretching $\nu_1$ state $(1,0,0)$, and the first and second excited bending $\nu_2$ state $(0,1^1,0)$ and $(0,2^0,0)$. Here, vibrational states are labelled with normal mode notation $(v_1,v_2^{L},v_3)$, where $v_1$ and $v_3$ correspond to the symmetric and asymmetric stretching modes, respectively, and $v_2$ labels the bending mode. The quantum number $L$ is related to the absolute value of the vibrational angular momentum quantum number $l$ associated with the $\nu_2$ bending mode, $L=|l|$. It is the $\nu_2$ bending mode which exhibits Renner-Teller splitting. No studies of the O--H stretching $\nu_3$ mode have been carried out. As a result, we do not allow potential parameters related to the $\nu_3$ mode to vary in the refinement and they are instead fixed to the original \textit{ab initio} values determined in \citet{jt838}. Computed energy levels and transitions involving the $\nu_3$ mode will therefore not be as accurate as those involving only the $\nu_1$ and $\nu_2$ modes in the final line list. This should not be overly problematic, however, as rovibronic transitions from the ground \X\ electronic state to the ground, $\nu_1$ and $\nu_2$ vibrational levels in the \A\ state are the strongest and these levels have all been characterised by experiment and utilised in the refinement.

The \A\ state PESs were refined using the \EVEREST\ code and the new implementation is discussed in Sec.~\ref{sec:extension_everest}. A total of 10 potential parameters of the $A^{\prime}$ and $A^{\prime\prime}$ surfaces were simultaneously varied in the refinement along with the zeroth-order parameter for the spin-orbit coupling surface, which was fine-tuned to ensure accurate spin-orbit splitting in the energy level structure of CaOH. 308 term values up to $J=15.5$, where $J$ is the total angular momentum quantum number, were used in the refinement and were reproduced with a root-mean-square (rms) error of 0.318~cm$^{-1}$. A weighting scheme based on the measurement uncertainty and number of transitions involving the energy level was utilised. This information was obtained from the CaOH MARVEL analysis~\citep{jt791}, essentially giving more importance in the refinement to well characterised term values with smaller uncertainties.

\begin{figure}
\centering
\includegraphics[width=0.49\textwidth]{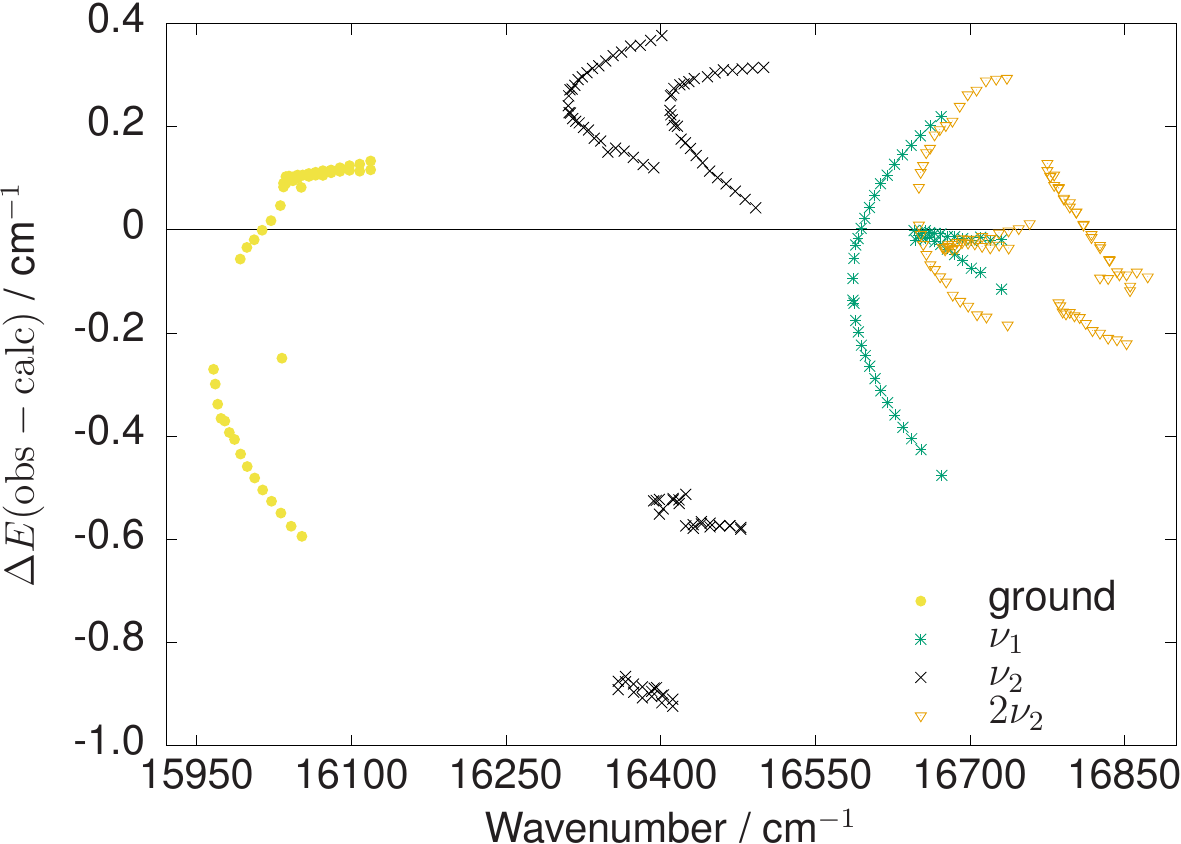}
\caption{\label{fig:res_pes_caoh}Residual errors $\Delta E(\mathrm{obs}-\mathrm{calc})$ between the empirically-derived MARVEL energy levels of CaOH and the calculated \EVEREST\ values using the refined potential energy surface for the different vibrational levels $v$ in the \A\ state.}
\end{figure}

The results of the \A\ PES refinement are illustrated in Fig.~\ref{fig:res_pes_caoh}, where we have plotted the residual errors $\Delta E({\rm obs-calc})$ between the empirically-derived MARVEL and calculated energies up to $J=15.5$. Generally speaking, the residual errors in each vibrational state exhibit similar behaviour with steadily increasing errors with increasing $J$ (or energy as seen in Fig.~\ref{fig:res_pes_caoh}) in both positive and negative directions. For example, in the ground vibrational state (yellow dots in Fig.~\ref{fig:res_pes_caoh}) the residual errors for the $e$-parity levels become more negative with $J$ (and energy), while the errors for the $f$-parity levels increase positively with $J$. There is also an outlier corresponding to the $J=1.5$, $f$-parity energy level at 16032.5597~cm$^{-1}$ and we suspect this to be an issue with the empirically-derived MARVEL value given the consistent behaviour of the residual errors for the other energy levels in the ground vibrational state.

Interestingly, the second excited \B\ state of CaOH has a minimum around 18\,000~cm$^{-1}$ and couples to the \A\ state through spin-orbit coupling and linear vibronic coupling. As we have neglected the \B\ state and its associated couplings in our spectroscopic model, there will no doubt be an impact on the accuracy that could be achieved had we been able to incorporate them. Even adding a ``dummy'' coupling curve as a free fitted parameter can improve the accuracy of the spectroscopic model as is commonly seen for diatomics studied by the ExoMol project, for example, see CaH~\citep{jt858}. The \A\ state is expected to acquire a significant fraction of \B\ character through the vibronic interaction~\citep{94CoLiPr.CaOH} and recently it was shown that the inclusion of the $\tilde{B}$--$\tilde{A}$ couplings can improve the calculation of vibronic branching ratios in CaOH~\citep{21ZhAuLa.CaOH}, which are needed when designing efficient laser-cooling schemes. That said, the effect of the \B\ state on the \A\ state is to some extent absorbed into the model by empirically refining the \A\ PESs.

\subsection{Extension of the \EVEREST\ code}
\label{sec:extension_everest}

The \EVEREST\ code~\citep{EVEREST} was used in all rotation-vibration (rovibrational) and rovibronic line list calculations and for the \A\ PES refinement. \EVEREST\ is capable of treating interacting electronic states with spin-dependent coupling and Renner-Teller effects in triatomic molecules. The program is general in its design and employs an exact kinetic energy operator. Extending the functionality of \EVEREST\ to be able to perform PES refinement required several developments.

The PESs of the  ground state $V_{X}$ and excited states $V_{A'}$, $V_{A''}$ are   represented using the Taylor-type expansion 
\begin{equation}
\label{eq:pot}
V_{S} =  \sum_{ijk} f_{ijk}^{(S)} \xi_1^{i} \xi_2^{j} \xi_3^{k}, \quad S = X, A', A'',
\end{equation}
where the coordinates
\begin{eqnarray}
\label{eq:coords_pes}
  \xi_1 &=& (r_1-r_1^{\rm eq})/r_1, \\
  \xi_2 &=& (r_2-r_2^{\rm eq})/r_2, \\
  \xi_3 &=& \alpha-\alpha_{\rm eq},
\end{eqnarray}
with the internal stretching coordinates $r_1  = r_{\rm CaO} $, $r_2  = r_{\rm OH} $, the interbond angle $\alpha = \angle({\rm CaOH})$, and the equilibrium parameters  $r_1^{\rm eq}$, $r_2^{\rm eq}$ and $\alpha_{\rm eq}$ (see \citet{jt838}).

The refinement procedure is a least-squares fitting of the potential parameters $f_{ijk}^{(S)}$ and structural parameters $r_i^{\rm eq}$ and $\alpha_{\rm eq}$ to the observed (MARVEL) energies of CaOH, realized through the minimization of the functional 
\begin{equation}
F = \sum_{i} w_i \left(E_{i}^{\rm (obs)} - E_{i}^{\rm (calc)}\right)^2,
\end{equation}
using Newton's steepest descent method. Here $E_{i}^{\rm (obs)}$ and $E_{i}^{\rm (calc)}$ are the observed and calculated rovibronic energies, respectively, and $w_i$ are fitting weights. The first derivatives of $E_{i} \equiv E_{i}^{\rm (calc)} $ with respect to (wrt) the varied potential parameters $f_{ijk}^{(S)}$ (required in the least-squares fitting) are evaluated using the Hellmann-Feynman theorem such that the derivatives of the rovibronic energies can be expressed as expectation values
\begin{equation}
\pdu{E_n}{f_{ijk}^{(S)}} = \bra{n}\pdu{\hat{H}}{f_{ijk}^{(S)}} \ket{n},
\end{equation}
where $\ket{n}$ is the corresponding rovibronic EVEREST wavefunction. This method takes advantage of the linear form of the $f_{ijk}^{(S)}$ parametrization,
\begin{equation}
  \bra{n}\pdu{\hat{H}}{f_{ijk}^{(S)}} \ket{n} = \bra{n} \xi_1^{i} \xi_2^{j} \xi_3^{k} \ket{n}.  
\end{equation}

Some of the $A'$ and $A''$ potential parameters of the two Renner-Taylor components corresponding to the linear geometry are shared between the two surfaces, namely the stretch-only terms $f_{ij0}^{(A')} = f_{ij0}^{(A'')} \equiv f_{ij0}^{(A)}$. The corresponding derivatives are given by the chain rule as follows,
\begin{equation}
\pdu{E_n}{f_{ij0}^{(A)}} = \pdu{E_n}{f_{ij0}^{(A')}} + \pdu{E_n}{f_{ij0}^{(A'')}}.  
\end{equation}
The derivatives wrt to the non-linear parameters $r_1^{\rm eq}$, $r_2^{\rm eq}$ and $\alpha_{\rm eq}$ are obtained using finite differences, for example,
\begin{equation}
\pdu{E_n}{r_1^{\rm eq}} = \frac{E_n(r_1^{\rm eq} + \Delta r)- E_n(r_1^{\rm eq} - \Delta r)}{2\Delta r},  
\end{equation}
where $\Delta r$ is a small displacement, typically taken as 0.1\% of the parameter value, 0.002~\AA\ or $0.2^{\circ}$ for the bond lengths or inter-bond angles, respectively. 

The implementation of the refinement procedure closely follows a previously developed Fortran wrapper for the triatomic nuclear motion program DVR3D reported in \citet{jt734}, where the underlying \textit{ab initio} PES was used to constrain the shape of the refined PES via a simultaneous fit to the original \textit{ab initio} energies. The fitting weights were initially set to be inversely proportional to the experimental (MARVEL) uncertainties and then adjusted via Watson's robust weighting scheme~\citep{03Watson.methods}. The expectation values of the potential expansions are now a part of the EVEREST program. The least-squares fitting procedure was implemented as an external, generalized fitting wrapper written in Fortran 95.

\subsection{\EVEREST\ line list calculations}

The final list list calculations in \EVEREST\ used the same set up and parameters as the PES refinement. Specifically, valence bond length-bond angle coordinates with a discrete variable representation (DVR) basis composed of 100 Sinc-DVR functions on both the Ca--O bond in the 2.6--7.0~$a_0$ interval and the O--H bond in the 1.1--6.0~$a_0$ interval, and with 120 Legendre functions for the $\angle({\rm CaOH})$ bond angle. The $J=0$ vibrational eigenfunctions were computed up to 10\,000~cm$^{-1}$ above the lowest vibronic state for $0\leq K\leq 27$, where $K=|\Lambda + l|$ ($\Lambda$ and $l$ are the projections of the electronic and vibrational angular momenta along the linear axis) from a Hamiltonian with a dimension of 10\,000. For $K\geq 1$, the Renner-Teller effect was explicitly taken into account by solving the coupled $A^{\prime}/A^{\prime\prime}$ problem, see \citet{EVEREST} for the full methodology. The full rovibronic Hamiltonian including spin-orbit coupling was built and diagonalized using these vibronic states for $J$ up to 175.5, where $J$ is the total angular momentum quantum number. Convergence of the computed rovibronic states was checked by running test calculations with $K\leq 50$, increasing the Hamiltonian dimension, and using larger DVR grids for the coordinates. The final selected \EVEREST\ calculation parameters produced sufficiently converged results and ensured that calculations were computationally tractable. All rovibrational transitions in the \X--\X\ band and rovibronic transitions in the \A--\X\ band were calculated to produce a line list containing 24,215,753,701 transitions between 3,187,522 states with rotational excitation up to $J=175.5$. The CaOH line list covers the 0\,--\,35\,000~cm$^{-1}$ range (wavelengths $\lambda > 0.29$~$\mu$m).

\section{Results}
\label{sec:results}

\subsection{Line list format}

The ExoMol data format has been discussed in detail~\citep{jt810} and is used for all the line lists in the ExoMol database. The \texttt{.trans} file, see Table~\ref{tab:trans}, contains the computed transitions with upper and lower state ID labels, Einstein $A$ coefficients (in s$^{-1}$) and transition wavenumbers (in cm$^{-1}$). The \texttt{.states} file, see Table~\ref{tab:states}, contains the computed rovibronic energy levels (in cm$^{-1}$), each labelled with a unique state ID counting number and quantum number labelling.

    To adapt the CaOH line list for high-resolution applications we have replaced the computed \EVEREST\ energy levels and their uncertainties with the empirical MARVEL values where available. This was done for a total of 1614 states up to $J=61.5$ for the ground \X\ and excited \A\ states. The final column in the \texttt{.states} file indicates whether the energy level is either calculated ({\tt Ca}) or from MARVEL ({\tt Ma}). All computed energy levels have an estimated uncertainty of 10~cm$^{-1}$, which is a conservative estimate to ensure that users are aware of the difference in reliability between calculated and ``MARVELised'' lines, particularly for high-resolution applications. The penultimate column of the \texttt{.states} file is the EVEREST calculated energy for reference. The quantum numbers used to label the rovibronic states of CaOH are listed in Table~\ref{tab:states} and discussed in detail in \citet{jt791}.

\begin{table}
\centering
\caption{\label{tab:trans}Extract from the \texttt{.trans} file of the CaOH OYT6 line list.}
\tt
\centering
\begin{tabular}{rrrr}
\toprule\toprule
\multicolumn{1}{c}{$f$}	&	\multicolumn{1}{c}{$i$}	& \multicolumn{1}{c}{$A_{fi}$}	&\multicolumn{1}{c}{$\tilde{\nu}_{fi}$} \\
\midrule
419 & 1 & 2.88286752E-08 & 0.6687\\
419 & 2 & 6.75986740E+00 & 353.3000\\
419 & 3 & 2.63388950E+00 & 609.7055\\
419 & 4 & 2.57185070E-02 & 689.3494\\
419 & 5 & 2.64598876E-01 & 951.0158\\
\bottomrule\bottomrule
\end{tabular} \\ \vspace{2mm}
\rm
\noindent
$f$: Upper  state counting number;\\
$i$:  Lower  state counting number; \\
$A_{fi}$:  Einstein-$A$ coefficient (in s$^{-1}$); \\
$\tilde{\nu}_{fi}$: Transition wavenumber (in cm$^{-1}$).\\
\end{table}

\begin{table*}
\centering
\caption{\label{tab:states} Extract from the \texttt{.states} file of the CaOH OYT6 line list. For the description of the quantum numbers see \citet{jt791}.}
{\setlength{\tabcolsep}{5pt}
\small\tt
\begin{tabular}{rr rrrrcrcrrrrrrrrr}
\toprule\toprule
$i$ & \multicolumn{1}{c}{$\tilde{E}$} (cm$^{-1}$) & $g_i$ & $J$ & \multicolumn{1}{c}{unc} & $\tau$ & $e/f$&$N$ & State & $L$ & $v_1$ & $v_2$ & $l_2$ & $v_3$ & $\Omega$ & $F_i$ & \multicolumn{1}{c}{Calc.} & Lab.  \\
 \midrule
        4056   &  16032.559740   &   8   &  1.5   &   0.014003 &  -1 &e  & 1& A' &  1 &   0 & 0 & 0 & 0 &  1.5 &F2 &  16032.218598 &  Ma \\
        4057   &  16311.011570   &   8   &  1.5   &   0.007071 &  -1 &e  & 1& A' &  0 &   0 & 1 & 1 & 0 &  0.5 &F1 &  16311.301542 &  Ma \\
        4058   &  16325.468962   &   8   &  1.5   &  10.000000 &  -1 &e  & 2& A' &  2 &   0 & 1 & 1 & 0 &  1.5 &F1 &  16325.468962 &  Ca \\
        4059   &  16410.656140   &   8   &  1.5   &   0.010000 &  -1 &e  & 2& A" &  0 &   0 & 1 & 1 & 0 &  0.5 &F1 &  16410.856088 &  Ma \\
        4060   &  16587.366240   &   8   &  1.5   &   0.005000 &  -1 &e  & 2& A' &  1 &   1 & 0 & 0 & 0 &  0.5 &F1 &  16587.401835 &  Ma \\
        4061   &  16645.654491   &   8   &  1.5   &  10.000000 &  -1 &e  & 1& A' &  1 &   1 & 0 & 0 & 0 &  1.5 &F2 &  16645.654491 &  Ca \\
        4062   &  16650.410240   &   8   &  1.5   &   0.010000 &  -1 &e  & 2& A' &  1 &   0 & 2 & 0 & 0 &  0.5 &F1 &  16650.559911 &  Ma \\
\bottomrule\bottomrule
\end{tabular}}
\mbox{}\\
{\flushleft
\begin{tabular}{ll}
\toprule\toprule
$i$:  &  State counting number.     \\
$\tilde{E}$: &  State energy (in cm$^{-1}$). \\
$g_i$: &  Total statistical weight, equal to ${g_{\rm ns}(2J + 1)}$.     \\
$J$: &  Total angular momentum.\\
unc: &  Uncertainty (in cm$^{-1}$).\\
$\tau$: &    Total parity. \\
$N$: &  Rotational angular momentum.\\
State: &  Electronic state X, A', A''.\\
$L$: &    Vibronic angular momentum quantum number, used to label vibronic states with $\Sigma$, $\Pi$, $\Delta$,  $\Phi$, \ldots for $L=0,1,2,3,\ldots$. \\ 
& $L=l_{\rm vib} + \Lambda$, where $l_{\rm vib}$ and  $\Lambda$ are  projections of the vibrational and  electronic angular momenta on the $z$ axis.    \\
$v_1$:  &  Symmetric stretching $\nu_1$ mode vibrational quantum number (``{\tt -1}'' means unassigned).\\
$v_2$: &   Bending $\nu_2$ mode vibrational quantum number (``{\tt -1}'' means unassigned).\\
$l_2$: &   Vibrational angular momentum quantum number associated with $\nu_2$ mode (``{\tt -1}'' means unassigned).\\
$v_3$: &   Antisymmetric stretching $\nu_3$ mode vibrational quantum number (``{\tt -1}'' means unassigned).\\
$\Omega$ &  Projection of the total angular momentum $|\Omega| \le J $.\\
$F_i$: &    Spin components $F_1$ and $F_3$.\\
Calc: &  Original \EVEREST\ calculated state energy (in cm$^{-1}$).\\
Label: & Label ``{\tt Ma}'' for MARVEL, ``{\tt Ca}'' for calculated. \\
\bottomrule
\end{tabular}
}
\end{table*}

\subsection{Temperature-dependent partition functions}
\label{sec:pfn}

\begin{figure}
\centering
\includegraphics[width=0.48\textwidth]{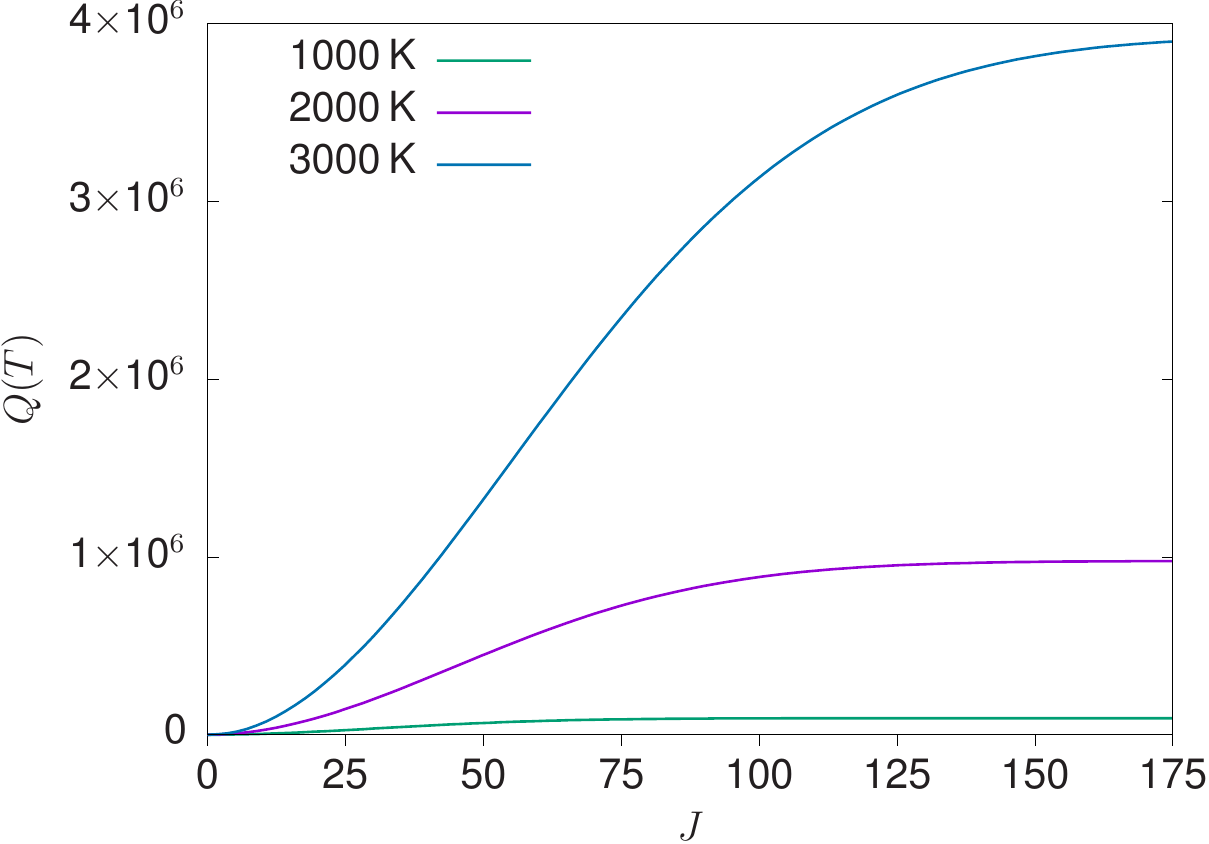}
\caption{\label{fig:pfn}Convergence of the partition function $Q(T)$ of CaOH with respect to the total angular momentum quantum number $J$ at different temperatures.}
\end{figure}

Intensity simulations require knowledge of the temperature-dependent partition function $Q(T)$, defined as
\begin{equation}
\label{eq:pfn}
Q(T)=\sum_{i} g_i \exp\left(\frac{-E_i}{kT}\right) ,
\end{equation}
where $g_i=g_{\rm ns}(2J_i+1)$ is the degeneracy of a state $i$ with energy $E_i$ and total angular momentum quantum number $J_i$, and the nuclear spin statistical weight $g_{\rm ns}=2$ for CaOH. Partition function values were computed on a $1$~K grid in the 1\,--\,4000~K range and can be downloaded from the ExoMol website at \href{www.exomol.com}{www.exomol.com} along with the OYT6 line list. The convergence of $Q(T)$ as a function of $J$ is shown in Fig.~\ref{fig:pfn} for different temperatures. At $T=3000$~K, the convergence of the partition function with $J$ is slower and we recommend this as a soft upper limit for using the OYT6 line list at elevated temperatures. Application of the OYT6 line list above this temperature may result in a progressive loss of opacity.

\begin{figure}
\centering
\includegraphics[width=0.48\textwidth]{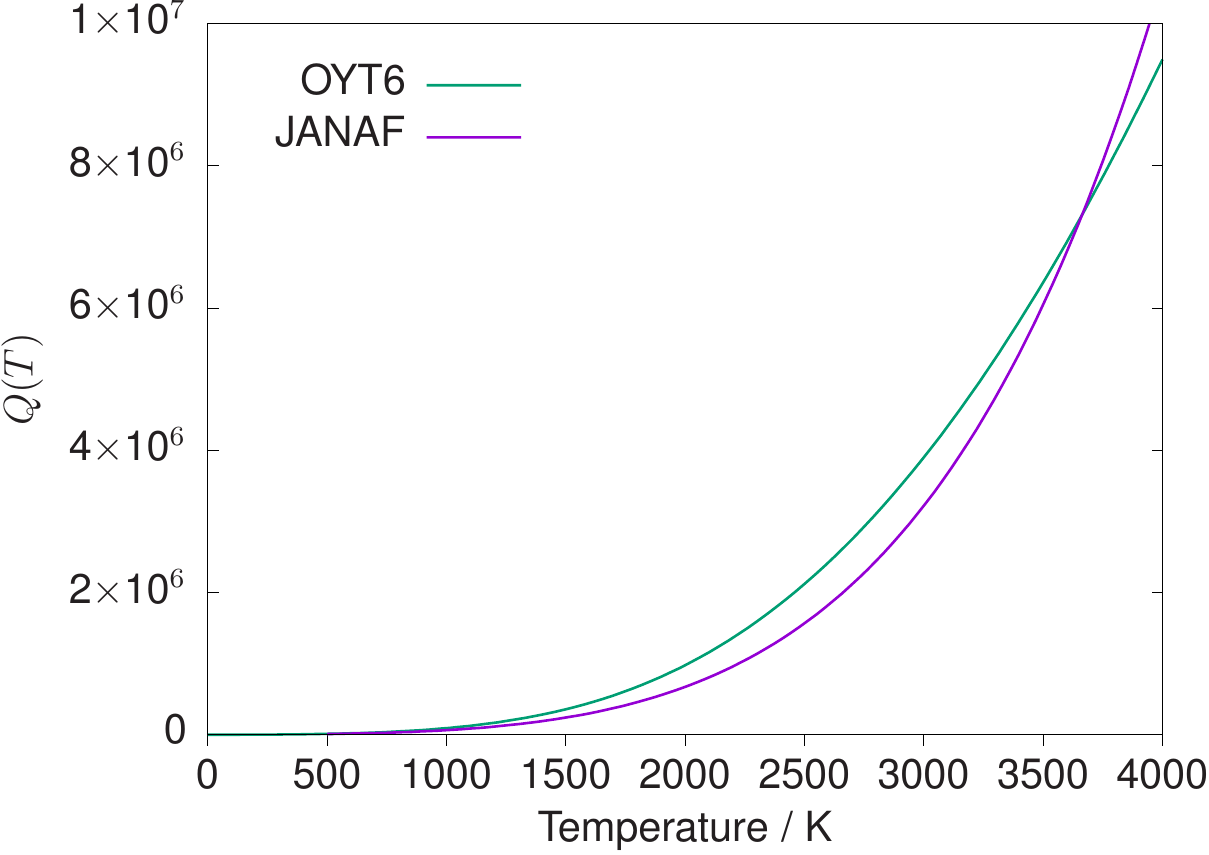}
\caption{\label{fig:pfn_janaf}Computed OYT6 partition function values $Q(T)$ of CaOH for temperatures up to $T=4000$~K compared against the JANAF values~\citep{88Irwin}.}
\end{figure}

In Fig.~\ref{fig:pfn_janaf} we have compared partition function values up to $T=4000$~K computed using the OYT6 line list against the JANAF CaOH values~\citep{88Irwin}. The JANAF values were calculated using a simple molecular model that employs a rigid rotor, simple harmonic oscillator approximation. As such, although the general behaviour between the OYT6 and JANAF values is in reasonable agreement, we expect the OYT6 values to be more accurate at least up to $T=3000$~K. This is because the OYT6 values are computed by summing over all meaningful rotation-vibration energy levels in the lowest two electronic states of CaOH.  Interestingly, the contribution from \A\ state energy levels in the summation of Eq.~\eqref{eq:pfn} to the partition function is near-negligible. For example, the partition function at $T=3000$~K calculated using only ground \X\ state energy levels in the summation of Eq.~\eqref{eq:pfn} showed a 0.08\% difference to the full OYT6 value and this is the largest difference encountered.

\subsection{Simulated spectra}

There is no meaningful intensity information in the literature on the $\tilde{A}$--$\tilde{X}$ band system that we can compare to, making it difficult to evaluate the accuracy of our computed line intensities against measured spectra. Transition intensities computed using \textit{ab initio} DMSs are comparable to, and in certain instances, more reliable than experiment~\citep{13Yurchenko.method,jt573}. As an estimate we would expect the CaOH line intensities to be within 5--10\% of experimentally determined values. It is also worth noting that strong electronic transitions are less sensitive to the shape of the transition DMSs, given that vibronic intensities can be relatively well modelled using only a single value for the transition dipole times a Franck-Condon factor. 

\begin{figure}
\centering
\includegraphics[width=0.7\textwidth]{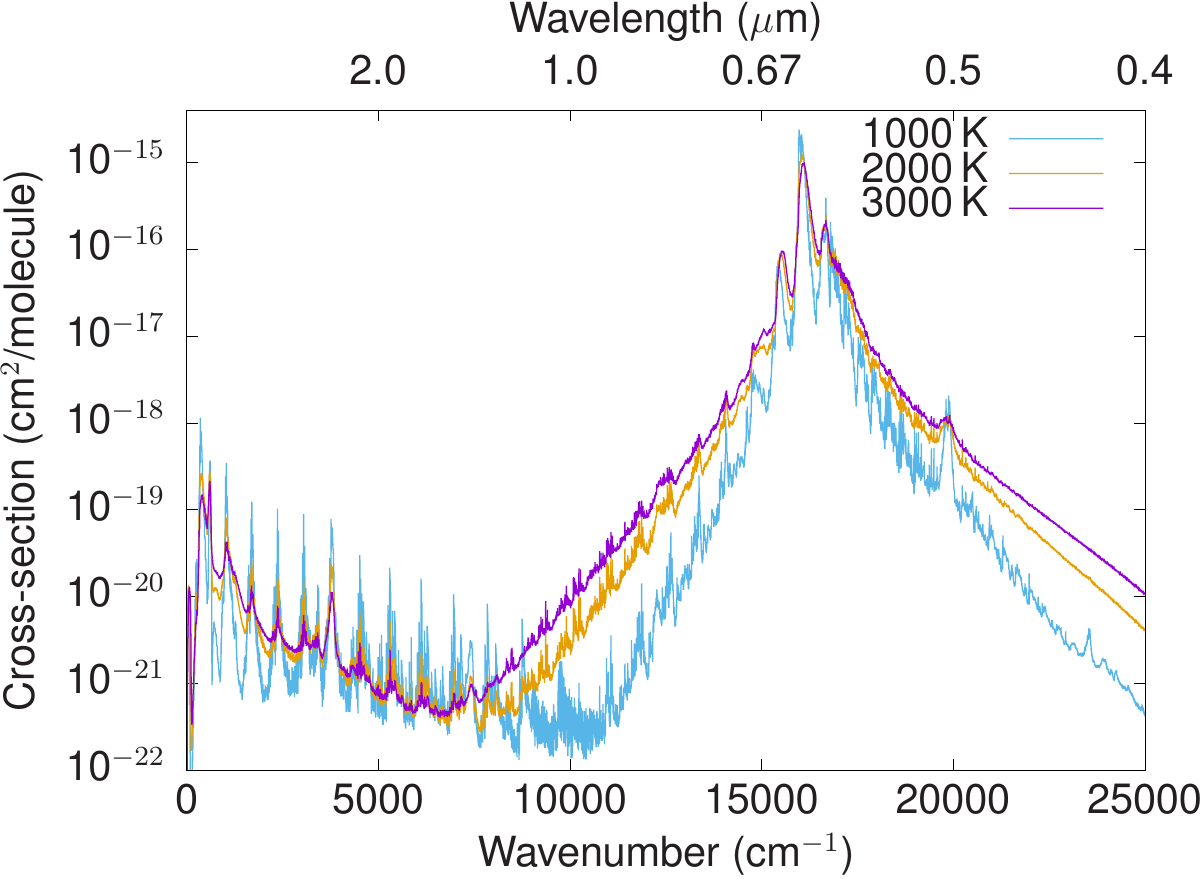}
\caption{\label{fig:1000K-3000K_caoh}Spectrum of CaOH at various temperatures (in K). Absorption cross-sections were computed at a resolution of 1~cm$^{-1}$ and modelled with a Gaussian line profile with a half width at half maximum (HWHM) of 1~cm$^{-1}$.}
\end{figure}

In Fig.~\ref{fig:1000K-3000K_caoh}, the temperature dependence of the CaOH spectrum is illustrated and we have simulated absolute absorption cross-sections at a resolution of 1~cm$^{-1}$ using a Gaussian line profile with a half width at half maximum (HWHM) of 1~cm$^{-1}$. All spectral simulations were done using the \textsc{ExoCross} program~\citep{jt708}. The strongest rovibronic features of the OYT6 line list are around 16\,000~cm$^{-1}$ and even at high temperatures ($T=3000$~K) these still dominate the CaOH spectrum. This region is significant for atmospheric studies of hot rocky super-Earths, a class of exoplanets that are in close proximity to their host star and exposed to extremely high temperatures, e.g.\ up to 4000~K.

\begin{figure}
\centering
\includegraphics[width=0.48\textwidth]{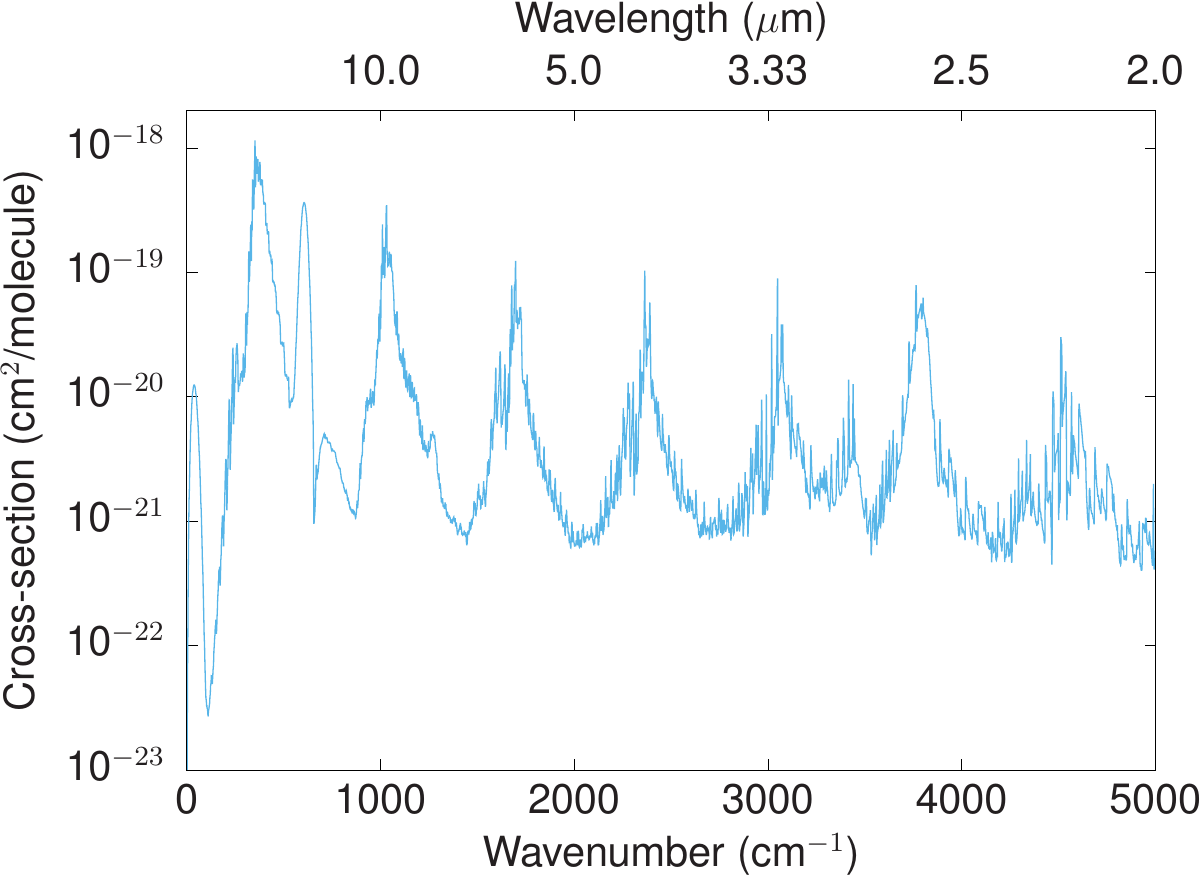}
\caption{\label{fig:ir_caoh}Absorption cross-sections at a temperature of $T=1000$~K in the 0--5000~cm$^{-1}$ range at a resolution of 1~cm$^{-1}$, modelled with a Gaussian line profile with a half width at half maximum (HWHM) of 1~cm$^{-1}$.}
\end{figure}

Closer inspection of the main infrared part of the spectrum is shown in Fig.~\ref{fig:ir_caoh} at $T=1000$~K. The fundamental wavenumbers of the \X\ state of CaOH occur at $\nu_1\approx 609$, $\nu_2\approx 353$ and $\nu_3\approx 3792$~cm$^{-1}$~\citep{jt838}. The spectrum exhibits a relatively slow flattening of band intensities with increasing wavenumber and similar behaviour was also observed in line list calculations of the alkali metal hydroxides KOH and NaOH \cite{jt820}, which both possess linear molecular structures.  

\begin{figure}
\centering
\includegraphics[width=0.49\textwidth]{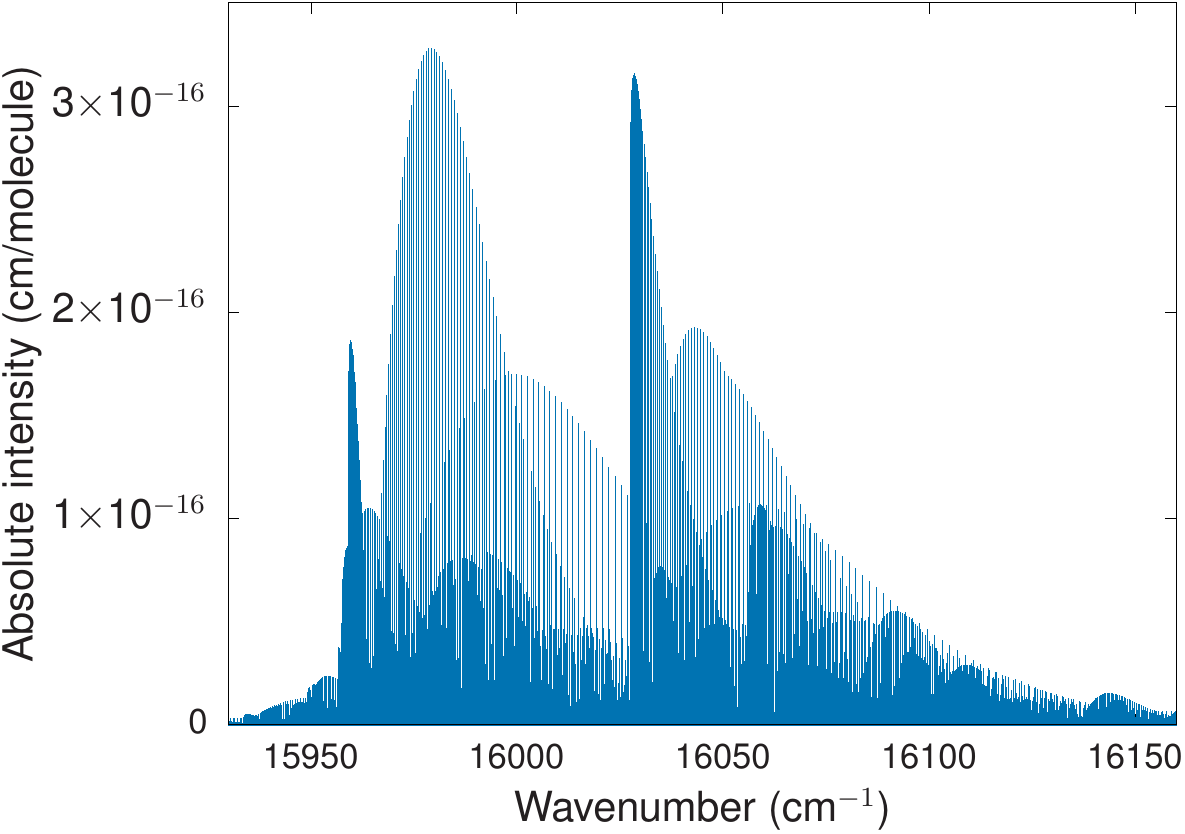}
\includegraphics[width=0.49\textwidth]{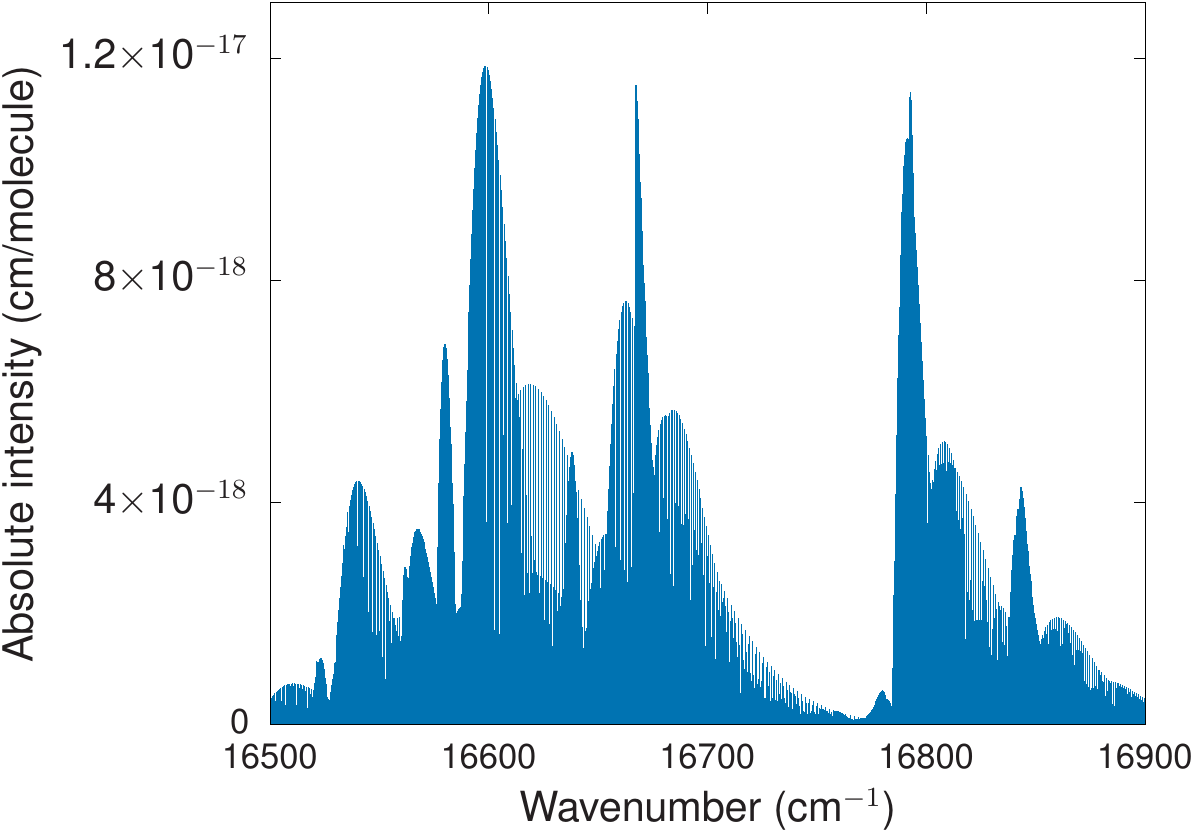}
\caption{\label{fig:caoh_stick}Absolute line intensities of the strongest \A\ bands of CaOH at a temperature of $T=700$~K.}
\end{figure}

In Fig.~\ref{fig:caoh_stick}, absolute absorption line intensities (in units of cm/molecule) at $T=700$~K have been plotted for the strongest regions of the OYT6 spectrum. Line intensities were simulated using the expression,
\begin{equation}
\label{eq:abs_I}
I(f \leftarrow i) = \frac{A_{fi}}{8\pi c}g_{\mathrm{ns}}(2 J_{f}+1)\frac{\exp\left(-E_{i}/kT\right)}{Q(T)\; \nu_{fi}^{2}}\left[1-\exp\left(-\frac{hc\nu_{fi}}{kT}\right)\right] ,
\end{equation}
where $A_{fi}$ is the Einstein $A$ coefficient of a transition with wavenumber $\nu_{fi}$ (in cm$^{-1}$) between an initial state with energy $E_i$ and a final state with rotational quantum number $J_f$. Here, $k$ is the Boltzmann constant, $h$ is the Planck constant, $c$ is the speed of light and $T$ is the absolute temperature. The nuclear spin statistical weight $g_{\mathrm{ns}}=2$ for CaOH and $Q(T)$ is the temperature-dependent partition function. Interestingly, the structure of the bands using the refined \A\ state PES are quantitatively different to those predicted using the original \textit{ab initio} PES, see Fig.~5 in \citet{jt838}. This behaviour has been encountered before by the ExoMol project and highlights the fact that rovibronic spectra can be very sensitive to the quality of the underlying PESs with both line positions and intensities being affected.

\begin{figure}
\centering
\includegraphics[width=0.48\textwidth]{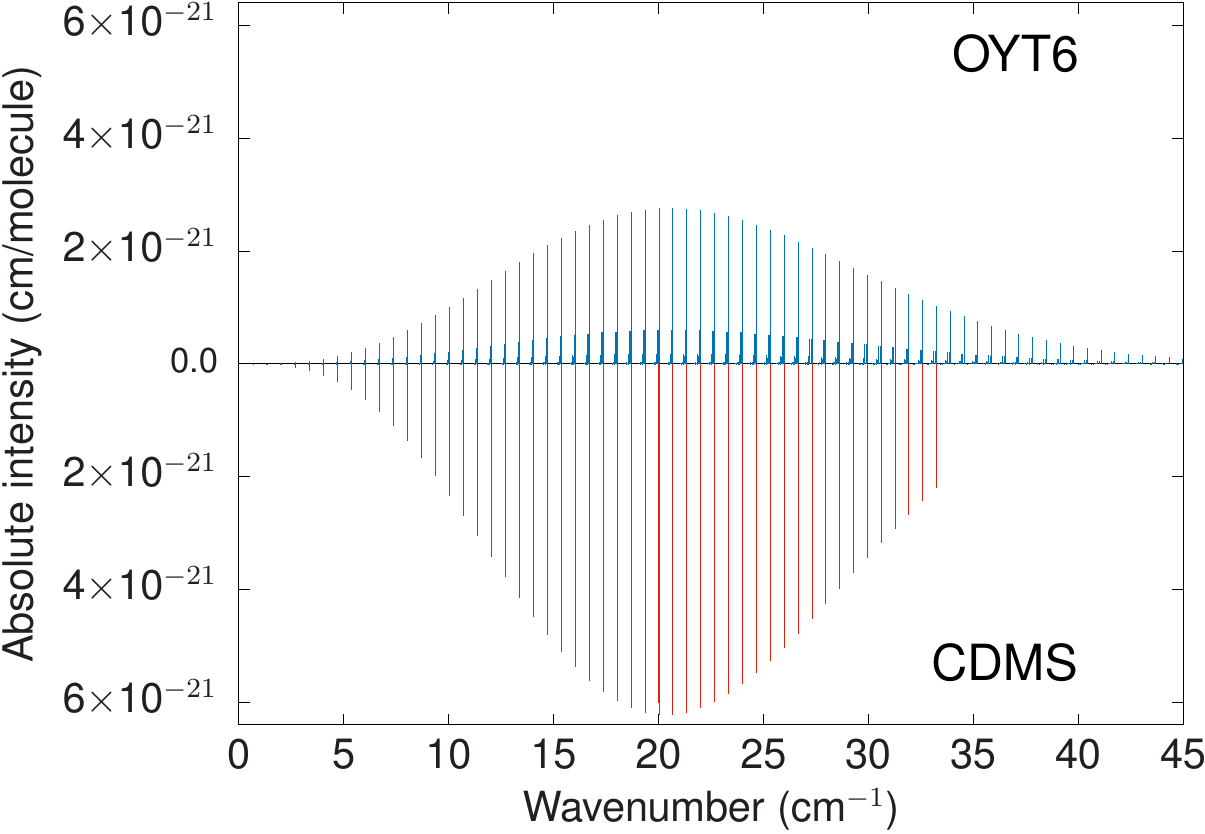}
\caption{\label{fig:rot_caoh}Comparison of the OYT6 stick spectrum at $T=300$~K in the microwave region against all transition data from the Cologne Database for Molecular Spectroscopy (CDMS)~\citep{CDMS:2001,CDMS:2005}.}
\end{figure}

In Fig.~\ref{fig:rot_caoh}, a CaOH OYT6 stick spectrum at $T=300$~K, computed using Eq.~\eqref{eq:abs_I}, is plotted against pure rotational transitions in the ground \X\ state extracted from the Cologne Database for Molecular Spectroscopy (CDMS)~\citep{CDMS:2001,CDMS:2005}. The rotational band is reproduced well with the OYT6 spectrum possessing more structure and lines since it extends up to $J=175.5$, compared to CDMS which contains 102 lines up to $J=50.5$. The CDMS line intensities are noticeably stronger, however, and they were computed using a more simplified model based on a single dipole moment value of $\mu=1.47$~Debye~\citep{92StFlJu.CaOH}. Our value for the ground \X\ state dipole is $\mu=1.09$~Debye and this was computed using high-level \textit{ab initio} theory (restricted coupled cluster with a large augmented correlation consistent basis set RCCSD(T)/aug-pwCVQZ-PP). This value is closer to other CaOH dipole values of 0.98~Debye from \textit{ab initio} calculations~\citep{90BaLaSt.CaOH} and a semiempirical value of 1.2~Debye~\citep{91MeVixx.CaOH}, and would explain the weaker OYT6 line intensities. These comparisons also give us confidence in the reliability of the DMSs used in intensity calculations for the OYT6 line list.

\section{Conclusion}
\label{sec:conc}

A new comprehensive molecular line list of the \X--\X\ rovibrational and \A--\X\ rovibronic band system of CaOH has been reported. The OYT6 line list covers rotational, rotation-vibration and the \A--\X\ electronic transition in the 0\,--\,35\,000~cm$^{-1}$ range (wavelengths $\lambda > 0.29$~$\mu$m) and is applicable to temperatures up to $T=3000$~K. For CaOH, we expect the strong \A--\X\ band around 16\,000~cm$^{-1}$ ($\lambda = 0.63$~$\mu$m) to be of significant interest as it dominates the spectrum at very high temperatures, which will be relevant to the conditions found on hot rocky super-Earth exoplanets. Also important is that the OYT6 line list has been adapted to high-resolution applications by replacing computed energy levels with the more accurate empirically-derived MARVEL values where available. To generate the OYT6 line list, extension of the computer program \EVEREST\ was necessary such that the \A\ state PESs of CaOH could be rigorously refined to experimental data, thus considerably improving the accuracy of the computed spectra. The extension of \EVEREST\ will also lend itself to future spectroscopic investigations, for example, the near-ultraviolet (UV) spectrum of water which is of considerable interest due to disagreement over absorption in this region, see \citet{22WaYiZh.H2O} and references within. Any future line list of H$_2$O in the near-UV region would find great applicability in exoplanetary research. 

\citet{13RaReAl.CaOH} used BT-Settl~\citep{BT-Settl} to analyse synthetic spectra of M-dwarf stars; they suggested that BT-Settl was missing key absorption due to three molecules: AlH and NaH in blue, and CaOH at 5570~\AA. Previous ExoMol studies have provided line lists and opacities for the metal hydrides AlH~\citep{jt732} and NaH~\citep{jt605}; provision of the OYT6 line list for CaOH means that the set should now be complete.

It is worth mentioning that the lowest-lying eight electronic states of CaOH up to the $\tilde{G}\,^2\Pi$ state~\citep{97HaJaBe.CaOH} at approximately 32\,633~cm$^{-1}$ are known. While it is unclear how strong these electronic transitions are in relation to the \A--\X\ band, the \B--\X\ band would almost certainly be observable as it lies in the same spectroscopic region (approximately 18\,000~cm$^{-1}$). Future work by the ExoMol database could address this electronic transition if deemed worthwhile.

\section*{Acknowledgments}

This work was supported by the STFC Projects No. ST/M001334/1 and ST/R000476/1. The authors acknowledge the use of the UCL Legion High Performance Computing Facility (Legion@UCL) and associated support services in the completion of this work, along with the Cambridge Service for Data Driven Discovery (CSD3), part of which is operated by the University of Cambridge Research Computing on behalf of the STFC DiRAC HPC Facility (www.dirac.ac.uk). The DiRAC component of CSD3 was funded by BEIS capital funding via STFC capital grants ST/P002307/1 and ST/R002452/1 and STFC operations grant ST/R00689X/1. DiRAC is part of the National e-Infrastructure. This work was also supported by the European Research Council (ERC) under the European Union’s Horizon 2020 research and innovation programme through Advance Grant number 883830.

\section*{Data Availability}

The states, transition and partition function files for the OYT6 line list can be downloaded from \href{www.exomol.com}{www.exomol.com} and the CDS data centre \href{http://cdsarc.u-strasbg.fr}{cdsarc.u-strasbg.fr}. The open access program \textsc{ExoCross} is available from \href{https://github.com/exomol}{github.com/exomol}.


\bibliographystyle{mnras}

\begin{thebibliography}{}
\makeatletter
\relax
\def\mn@urlcharsother{\let\do\@makeother \do\$\do\&\do\#\do\^\do\_\do\%\do\~}
\def\mn@doi{\begingroup\mn@urlcharsother \@ifnextchar [ {\mn@doi@}
  {\mn@doi@[]}}
\def\mn@doi@[#1]#2{\def\@tempa{#1}\ifx\@tempa\@empty \href
  {http://dx.doi.org/#2} {doi:#2}\else \href {http://dx.doi.org/#2} {#1}\fi
  \endgroup}
\def\mn@eprint#1#2{\mn@eprint@#1:#2::\@nil}
\def\mn@eprint@arXiv#1{\href {http://arxiv.org/abs/#1} {{\tt arXiv:#1}}}
\def\mn@eprint@dblp#1{\href {http://dblp.uni-trier.de/rec/bibtex/#1.xml}
  {dblp:#1}}
\def\mn@eprint@#1:#2:#3:#4\@nil{\def\@tempa {#1}\def\@tempb {#2}\def\@tempc
  {#3}\ifx \@tempc \@empty \let \@tempc \@tempb \let \@tempb \@tempa \fi \ifx
  \@tempb \@empty \def\@tempb {arXiv}\fi \@ifundefined
  {mn@eprint@\@tempb}{\@tempb:\@tempc}{\expandafter \expandafter \csname
  mn@eprint@\@tempb\endcsname \expandafter{\@tempc}}}

\bibitem[\protect\citeauthoryear{{Allard}}{{Allard}}{2014}]{BT-Settl}
{Allard} F.,  2014, in {Booth} M.,  {Matthews} B.~C.,   {Graham} J.~R.,  eds,
  IAU Symposium Vol. 299, IAU Symposium. pp 271--272,
  \mn@doi{10.1017/S1743921313008545}

\bibitem[\protect\citeauthoryear{Augustovi{\v{c}}ov{\'{a}} \&
  Bohn}{Augustovi{\v{c}}ov{\'{a}} \& Bohn}{2019}]{19AuBoxx.CaOH}
Augustovi{\v{c}}ov{\'{a}} L.~D.,  Bohn J.~L.,  2019, \mn@doi [New J. Phys]
  {10.1088/1367-2630/ab4720}, 21, 103022

\bibitem[\protect\citeauthoryear{Baum, Vilas, Hallas, Augenbraun, Raval, Mitra
  \& Doyle}{Baum et~al.}{2020}]{20BaViHa.CaOH}
Baum L.,  Vilas N.~B.,  Hallas C.,  Augenbraun B.~L.,  Raval S.,  Mitra D.,
  Doyle J.~M.,  2020, \mn@doi [Phys. Rev. Lett.]
  {10.1103/physrevlett.124.133201}, 124, 133201

\bibitem[\protect\citeauthoryear{Baum, Vilas, Hallas, Augenbraun, Raval, Mitra
  \& Doyle}{Baum et~al.}{2021}]{21BaViHa.CaOH}
Baum L.,  Vilas N.~B.,  Hallas C.,  Augenbraun B.~L.,  Raval S.,  Mitra D.,
  Doyle J.~M.,  2021, \mn@doi [Phys. Rev. A] {10.1103/physreva.103.043111},
  103, 043111

\bibitem[\protect\citeauthoryear{Bauschlicher, Langhoff, Steimle  \&
  Shirley}{Bauschlicher et~al.}{1990}]{90BaLaSt.CaOH}
Bauschlicher C.~W.,  Langhoff S.~R.,  Steimle T.~C.,   Shirley J.~E.,  1990,
  \mn@doi [J. Chem. Phys.] {10.1063/1.459689}, 93, 4179

\bibitem[\protect\citeauthoryear{Bernath}{Bernath}{2009}]{09Bernath.exo}
Bernath P.~F.,  2009, \mn@doi [Int. Rev. Phys. Chem.]
  {10.1080/01442350903292442}, 28, 681

\bibitem[\protect\citeauthoryear{Bernath \& Brazier}{Bernath \&
  Brazier}{1985}]{85BeBrxx.CaOH}
Bernath P.~F.,  Brazier C.~R.,  1985, \mn@doi [ApJ] {10.1086/162800}, 288, 373

\bibitem[\protect\citeauthoryear{Birkby}{Birkby}{2018}]{18Birkby}
Birkby J.~L.,  2018, Handbook of Exoplanets, pp 1485--1508

\bibitem[\protect\citeauthoryear{Bowesman, Shuai, Yurchenko  \&
  Tennyson}{Bowesman et~al.}{2021}]{jt835}
Bowesman C.~A.,  Shuai M.,  Yurchenko S.~N.,   Tennyson J.,  2021, \mn@doi
  [MNRAS] {10.1093/mnras/stab2525}, 508, 3181

\bibitem[\protect\citeauthoryear{Coxon, Li  \& Presunka}{Coxon
  et~al.}{1991}]{91CoLiPr.CaOH}
Coxon J.~A.,  Li M.~G.,   Presunka P.~I.,  1991, \mn@doi [J. Mol. Spectrosc.]
  {10.1016/0022-2852(91)90191-C}, 150, 33

\bibitem[\protect\citeauthoryear{Coxon, Li  \& Presunka}{Coxon
  et~al.}{1992}]{92CoLiPr.CaOH}
Coxon J.~A.,  Li M.~G.,   Presunka P.~I.,  1992, \mn@doi [Mol. Phys.]
  {10.1080/00268979200102231}, 76, 1463

\bibitem[\protect\citeauthoryear{Coxon, Li  \& Presunka}{Coxon
  et~al.}{1994}]{94CoLiPr.CaOH}
Coxon J.~A.,  Li M.~G.,   Presunka P.~I.,  1994, \mn@doi [J. Mol. Spectrosc.]
  {10.1006/jmsp.1994.1060}, 164, 118

\bibitem[\protect\citeauthoryear{Cs{\'a}sz{\'a}r, Czak{\'o}, Furtenbacher  \&
  M{\'a}tyus}{Cs{\'a}sz{\'a}r et~al.}{2007}]{07CsCzFu.method}
Cs{\'a}sz{\'a}r A.~G.,  Czak{\'o} G.,  Furtenbacher T.,   M{\'a}tyus E.,  2007,
  \mn@doi [Annu. Rep. Comput. Chem.] {10.1016/s1574-1400(07)03009-5}, 3, 155

\bibitem[\protect\citeauthoryear{Dick, Sheridan, Wang, Yu  \& Bernath}{Dick
  et~al.}{2006}]{06DiShWa.CaOH}
Dick M.~J.,  Sheridan P.~M.,  Wang J.~G.,  Yu S.,   Bernath P.~F.,  2006,
  \mn@doi [J. Mol. Spectrosc.] {10.1016/j.jms.2006.10.009}, 240, 238

\bibitem[\protect\citeauthoryear{Furtenbacher \& {Cs\'asz\'ar}}{Furtenbacher \&
  {Cs\'asz\'ar}}{2012}]{12FuCsxx.methods}
Furtenbacher T.,  {Cs\'asz\'ar} A.~G.,  2012, \mn@doi [J. Mol. Struct.]
  {10.1016/j.molstruc.2011.10.057}, 1009, 123

\bibitem[\protect\citeauthoryear{Furtenbacher, {Cs\'asz\'ar}  \&
  Tennyson}{Furtenbacher et~al.}{2007}]{jt412}
Furtenbacher T.,  {Cs\'asz\'ar} A.~G.,   Tennyson J.,  2007, \mn@doi [J. Mol.
  Spectrosc.] {10.1016/j.jms.2007.07.005}, 245, 115

\bibitem[\protect\citeauthoryear{Hailey, Jarman  \& Bernath}{Hailey
  et~al.}{1997}]{97HaJaBe.CaOH}
Hailey R.~A.,  Jarman C.,   Bernath P.~F.,  1997, \mn@doi [J. Chem. Phys.]
  {10.1063/1.474428}, 107, 669

\bibitem[\protect\citeauthoryear{Hilborn, Qingshi  \& Harris}{Hilborn
  et~al.}{1983}]{83HiQiHa.CaOH}
Hilborn R.~C.,  Qingshi Z.,   Harris D.~O.,  1983, \mn@doi [J. Mol. Spectrosc.]
  {10.1016/0022-2852(83)90338-7}, 97, 73

\bibitem[\protect\citeauthoryear{{Irwin}}{{Irwin}}{1988}]{88Irwin}
{Irwin} A.~W.,  1988, A\&AS, \href
  {https://ui.adsabs.harvard.edu/abs/1988A&AS...74..145I} {74, 145}

\bibitem[\protect\citeauthoryear{Jarman \& Bernath}{Jarman \&
  Bernath}{1992}]{92JaBexx.CaOH}
Jarman C.~N.,  Bernath P.~F.,  1992, \mn@doi [J. Chem. Phys.]
  {10.1063/1.463158}, 97, 1711

\bibitem[\protect\citeauthoryear{Kozyryev, Steimle, Yu, Nguyen  \&
  Doyle}{Kozyryev et~al.}{2019}]{19KoStYu.CaOH}
Kozyryev I.,  Steimle T.~C.,  Yu P.,  Nguyen D.-T.,   Doyle J.~M.,  2019,
  \mn@doi [New J. Phys.] {10.1088/1367-2630/ab19d7}, 21, 052002

\bibitem[\protect\citeauthoryear{Li \& Coxon}{Li \&
  Coxon}{1992}]{92LiCoxx.CaOH}
Li M.~G.,  Coxon J.~A.,  1992, \mn@doi [J. Chem. Phys.] {10.1063/1.463322}, 97,
  8961

\bibitem[\protect\citeauthoryear{Li \& Coxon}{Li \&
  Coxon}{1995}]{95LiCoxx.CaOH}
Li M.~G.,  Coxon J.~A.,  1995, \mn@doi [J. Chem. Phys.] {10.1063/1.468643},
  102, 2663

\bibitem[\protect\citeauthoryear{Li \& Coxon}{Li \&
  Coxon}{1996}]{96LiCoxx.CaOH}
Li M.~G.,  Coxon J.~A.,  1996, \mn@doi [J. Chem. Phys.] {10.1063/1.471762},
  104, 4961

\bibitem[\protect\citeauthoryear{Mestdagh \& Visticot}{Mestdagh \&
  Visticot}{1991}]{91MeVixx.CaOH}
Mestdagh J.~M.,  Visticot J.~P.,  1991, \mn@doi [Chem. Phys.]
  {10.1016/0301-0104(91)87008-J}, 155, 79

\bibitem[\protect\citeauthoryear{Mitrushchenkov}{Mitrushchenkov}{2012}]{EVEREST}
Mitrushchenkov A.~O.,  2012, \mn@doi [J. Chem. Phys.] {10.1063/1.3672162}, 136,
  024108

\bibitem[\protect\citeauthoryear{M\"{u}ller, Thorwirth, Roth  \&
  Winnewisser}{M\"{u}ller et~al.}{2001}]{CDMS:2001}
M\"{u}ller H. S.~P.,  Thorwirth S.,  Roth D.~A.,   Winnewisser G.,  2001,
  \mn@doi [A\&A] {10.1051/0004-6361:20010367}, 370, L49

\bibitem[\protect\citeauthoryear{{M\"uller}, {Schl\"oder}, Stutzki  \&
  Winnewisser}{{M\"uller} et~al.}{2005}]{CDMS:2005}
{M\"uller} H. S.~P.,  {Schl\"oder} F.,  Stutzki J.,   Winnewisser G.,  2005,
  \mn@doi [J. Mol. Struct.] {10.1016/j.molstruc.2005.01.027}, 742, 215

\bibitem[\protect\citeauthoryear{Owens, Zak, Chubb, Yurchenko, Tennyson  \&
  Yachmenev}{Owens et~al.}{2017}]{jt663}
Owens A.,  Zak E.~J.,  Chubb K.~L.,  Yurchenko S.~N.,  Tennyson J.,   Yachmenev
  A.,  2017, \mn@doi [Sci. Rep.] {10.1038/srep45068}, 45068, 7

\bibitem[\protect\citeauthoryear{Owens, Clark, Mitrushchenkov, Yurchenko  \&
  Tennyson}{Owens et~al.}{2021a}]{jt838}
Owens A.,  Clark V. H.~J.,  Mitrushchenkov A.,  Yurchenko S.~N.,   Tennyson J.,
   2021a, \mn@doi [J. Chem. Phys.] {10.1063/5.0052958}, 154, 234302

\bibitem[\protect\citeauthoryear{Owens, Tennyson  \& Yurchenko}{Owens
  et~al.}{2021b}]{jt820}
Owens A.,  Tennyson J.,   Yurchenko S.~N.,  2021b, \mn@doi [MNRAS]
  {10.1093/mnras/staa4041}, 502, 1128

\bibitem[\protect\citeauthoryear{Owens, Dooley, McLaughlin, Tan, Zhang,
  Yurchenko  \& Tennyson}{Owens et~al.}{2022}]{jt858}
Owens A.,  Dooley S.,  McLaughlin L.,  Tan B.,  Zhang G.,  Yurchenko S.~N.,
  Tennyson J.,  2022, \mn@doi [MNRAS] {10.1093/mnras/stac371}, 511, 5448

\bibitem[\protect\citeauthoryear{Polyansky, Kyuberis, Zobov, Tennyson,
  Yurchenko  \& Lodi}{Polyansky et~al.}{2018}]{jt734}
Polyansky O.~L.,  Kyuberis A.~A.,  Zobov N.~F.,  Tennyson J.,  Yurchenko S.~N.,
    Lodi L.,  2018, \mn@doi [MNRAS] {10.1093/mnras/sty1877}, 480, 2597

\bibitem[\protect\citeauthoryear{Rajpurohit, Reyle, Allard, Homeier,
  Schultheis, Bessell  \& Robin}{Rajpurohit et~al.}{2013}]{13RaReAl.CaOH}
Rajpurohit A.~S.,  Reyle C.,  Allard F.,  Homeier D.,  Schultheis M.,  Bessell
  M.~S.,   Robin A.~C.,  2013, \mn@doi [A\&A] {10.1051/0004-6361/201321346},
  556, A15

\bibitem[\protect\citeauthoryear{Renner}{Renner}{1934}]{Renner1934}
Renner R.,  1934, \mn@doi [Z. Phys.] {10.1007/bf01350054}, 92, 172

\bibitem[\protect\citeauthoryear{Rivlin, Lodi, Yurchenko, Tennyson  \& {Le
  Roy}}{Rivlin et~al.}{2015}]{jt605}
Rivlin T.,  Lodi L.,  Yurchenko S.~N.,  Tennyson J.,   {Le Roy} R.~J.,  2015,
  \mn@doi [MNRAS] {10.1093/mnras/stv979}, 451, 5153

\bibitem[\protect\citeauthoryear{Snellen}{Snellen}{2014}]{14Snellen}
Snellen I.,  2014, \mn@doi [Phil. Trans. Royal Soc. London A]
  {10.1098/rsta.2013.0075}, 372, 20130075

\bibitem[\protect\citeauthoryear{Steimle, Fletcher, Jung  \& Scurlock}{Steimle
  et~al.}{1992}]{92StFlJu.CaOH}
Steimle T.~C.,  Fletcher D.~A.,  Jung K.~Y.,   Scurlock C.~T.,  1992, \mn@doi
  [J. Chem. Phys.] {10.1063/1.462007}, 96, 2556

\bibitem[\protect\citeauthoryear{Tennyson}{Tennyson}{2014}]{jt573}
Tennyson J.,  2014, \mn@doi [J. Mol. Spectrosc.] {10.1016/j.jms.2014.01.012},
  298, 1

\bibitem[\protect\citeauthoryear{Tennyson \& Yurchenko}{Tennyson \&
  Yurchenko}{2012}]{jt528}
Tennyson J.,  Yurchenko S.~N.,  2012, \mn@doi [MNRAS]
  {10.1111/j.1365-2966.2012.21440.x}, 425, 21

\bibitem[\protect\citeauthoryear{Tennyson \& Yurchenko}{Tennyson \&
  Yurchenko}{2017}]{jt693}
Tennyson J.,  Yurchenko S.~N.,  2017, \mn@doi [Mol. Astrophys.]
  {10.1016/j.molap.2017.05.002}, 8, 1

\bibitem[\protect\citeauthoryear{Tennyson et~al.,}{Tennyson
  et~al.}{2016}]{jt631}
Tennyson J.,  et~al., 2016, \mn@doi [J. Mol. Spectrosc.]
  {10.1016/j.jms.2016.05.002}, 327, 73

\bibitem[\protect\citeauthoryear{Tennyson et~al.,}{Tennyson
  et~al.}{2020}]{jt810}
Tennyson J.,  et~al., 2020, \mn@doi [J. Quant. Spectrosc. Radiat. Transf.]
  {10.1016/j.jqsrt.2020.107228}, 255, 107228

\bibitem[\protect\citeauthoryear{T\'obi\'as, Furtenbacher, Tennyson  \&
  Cs\'asz\'ar}{T\'obi\'as et~al.}{2019}]{jt750}
T\'obi\'as R.,  Furtenbacher T.,  Tennyson J.,   Cs\'asz\'ar A.~G.,  2019,
  \mn@doi [Phys. Chem. Chem. Phys.] {10.1039/c8cp05169k}, 21, 3473

\bibitem[\protect\citeauthoryear{Vilas, Hallas, Anderegg, Robichaud, Winnicki,
  Mitra  \& Doyle}{Vilas et~al.}{2022}]{22ViHaAn.CaOH}
Vilas N.~B.,  Hallas C.,  Anderegg L.,  Robichaud P.,  Winnicki A.,  Mitra D.,
   Doyle J.~M.,  2022, \mn@doi [Nature] {10.1038/s41586-022-04620-5}, 606, 70

\bibitem[\protect\citeauthoryear{Wang, Owens, Tennyson  \& Yurchenko}{Wang
  et~al.}{2020}]{jt791}
Wang Y.,  Owens A.,  Tennyson J.,   Yurchenko S.~N.,  2020, \mn@doi [ApJS]
  {10.3847/1538-4365/ab85cb}, 248, 9

\bibitem[\protect\citeauthoryear{Wang, Yin, Min  \& Zhu}{Wang
  et~al.}{2022}]{22WaYiZh.H2O}
Wang Z.-C.,  Yin B.,  Min Q.,   Zhu L.,  2022, \mn@doi [J. Quant. Spectrosc.
  Radiat. Transf.] {10.1016/j.jqsrt.2022.108204}, 286, 108204

\bibitem[\protect\citeauthoryear{Watson}{Watson}{2003}]{03Watson.methods}
Watson J. K.~G.,  2003, \mn@doi [J. Mol. Spectrosc.]
  {10.1016/S0022-2852(03)00100-0}, 219, 326

\bibitem[\protect\citeauthoryear{Yurchenko}{Yurchenko}{2014}]{13Yurchenko.method}
Yurchenko S.~N.,  2014, in , Vol.~10, Chemical Modelling: Volume 10.
The Royal Society of Chemistry, Chapt.~7, pp 183--228,
  \mn@doi{10.1039/9781849737241-00183}

\bibitem[\protect\citeauthoryear{Yurchenko, Williams, Leyland, Lodi  \&
  Tennyson}{Yurchenko et~al.}{2018a}]{jt732}
Yurchenko S.~N.,  Williams H.,  Leyland P.~C.,  Lodi L.,   Tennyson J.,  2018a,
  \mn@doi [MNRAS] {10.1093/mnras/sty1524}, 479, 1401

\bibitem[\protect\citeauthoryear{Yurchenko, Al-Refaie  \& Tennyson}{Yurchenko
  et~al.}{2018b}]{jt708}
Yurchenko S.~N.,  Al-Refaie A.~F.,   Tennyson J.,  2018b, \mn@doi [A\&A]
  {10.1051/0004-6361/201732531}, 614, A131

\bibitem[\protect\citeauthoryear{Zhang, Augenbraun, Lasner, Vilas, Doyle  \&
  Cheng}{Zhang et~al.}{2021}]{21ZhAuLa.CaOH}
Zhang C.,  Augenbraun B.~L.,  Lasner Z.~D.,  Vilas N.~B.,  Doyle J.~M.,   Cheng
  L.,  2021, \mn@doi [J. Chem. Phys.] {10.1063/5.0063611}, 155, 091101

\makeatother
\end{thebibliography}

\bsp	
\label{lastpage}
\end{document}